\documentclass[a4paper,11pt]{article}
\usepackage{jheppub} 
\usepackage{lineno}

\arxivnumber{2412.19369} 

\usepackage{slashed}
\usepackage{feynmp-auto}
\usepackage{dcolumn}
\usepackage{bm}
\usepackage{epsfig}
\usepackage{float}
\usepackage{graphicx}

\usepackage[normalem]{ulem}

\usepackage{blkarray,booktabs}

\usepackage{mathrsfs}
\usepackage{color}
\usepackage[usenames,dvipsnames]{xcolor}
\usepackage{hyperref}

\def \cre{\color{red}}

\usepackage[caption=false]{subfig}

\usepackage{verbatim}

\def\fun#1#2{\lower3.6pt\vbox{\baselineskip0pt\lineskip.9pt
		\ialign{$\mathsurround=0pt#1\hfil##\hfil$\crcr#2\crcr\sim\crcr}}}

\input epsf

\newenvironment{Eqnarray}%
{\arraycolsep 0.14em\begin{eqnarray}}{\end{eqnarray}}
\newcommand{\be}{\begin{equation}}
	\newcommand{\ee}{\end{equation}}
\newcommand{\bea}{\begin{Eqnarray}}
	\newcommand{\eea}{\end{Eqnarray}}

\def\lsim{\mathrel{\raise.3ex\hbox{$<$\kern-.75em\lower1ex\hbox{$\sim$}}}}
\def\gsim{\mathrel{\raise.3ex\hbox{$>$\kern-.75em\lower1ex\hbox{$\sim$}}}}
\def\lsub#1{_{\lower 1.5pt\hbox{$\scriptstyle#1$}}}

\def\be{\begin{equation}}
	\def\ee{\end{equation}}
\def\bea{\begin{eqnarray}}
	\def\eea{\end{eqnarray}}

\begin{document}
\preprint{WSU-HEP-2406}
\title{
{\bf The Deconstruction of Flavor in the Privately Democratic Higgs Sector}}
\author{Bhubanjyoti Bhattacharya,}
\emailAdd{bbhattach@ltu.edu}
\affiliation{Department of Natural Sciences, Lawrence Technological University, Southfield, MI 48075, USA}

\author{Suneth Jayawardana,}
\emailAdd{sunethjayawardana@wayne.edu}
\affiliation{Department of Physics and Astronomy, Wayne State University, Detroit, MI 48201, USA}

\author{Nausheen R. Shah}
\emailAdd{nausheen.shah@wayne.edu}
\affiliation{Department of Physics and Astronomy, Wayne State University, Detroit, MI 48201, USA}

\abstract{
	The Standard Model (SM) of particle physics fails to explain the observed hierarchy in fermion masses or the origin of fermion-flavor structure. We construct a model to explain these observations in the quark sector. We introduce a spectrum of new particles consisting of six of each -- massive singlet vector-like quarks (VLQs), singlet scalars, and SU(2)-doublet scalars. SM quark masses are generated when the neutral components of the SU(2)-doublet scalars acquire non-zero vacuum expectation values (VEVs). We impose global symmetries to ensure that Yukawa couplings stay roughly flavor diagonal and democratic (of the same order), as well as to suppress tree-level flavor-changing neutral currents. Quark-mass hierarchy then follows from a hierarchy in scalar VEVs. The singlet scalars also acquire  weak-scale VEVs. Together with the VLQs, they act as messengers between different  generations of quarks in the SM. These messenger particles are responsible for generating the elements of the Cabibbo-Kobayashi-Masakawa (CKM) matrix which depend on the ratios of the singlet VEVs and VLQ masses. Constructed this way, the CKM matrix is found to be \emph{independent} of the SM fermion masses. Using the measured values of the CKM matrix elements and assuming order-one couplings, we derive constraints on the masses of the VLQs and discuss prospects for probing our model in the near future.
}

\maketitle

\section{Introduction}\label{sec:1}
The Standard Model (SM) of particle physics contains six quark flavors with widely different masses generated via the Higgs mechanism. Electroweak symmetry breaking (EWSB) \cite{Pich:2015lkh} occurs once the neutral Higgs field acquires a non-zero vacuum expectation value (VEV) and leads to massive weak gauge bosons. In addition, the Yukawa interactions between the Higgs field and SM quarks lead to quark masses proportional to the Higgs VEV: $m_q = y_q~v/\sqrt{2}$, where $m_q$ stands for the mass of a quark, $y_q$ its Yukawa coupling, and $v=246$ GeV the Higgs VEV. Since the Higgs VEV is universal to all quarks, the large hierarchy in quark masses -- a light up quark ($m_u \sim 3$ MeV) compared to a very heavy top quark ($m_t \sim 170$ GeV) -- can be attributed to a hierarchy in the dimensionless Yukawa couplings: $y_t/y_u = m_t/m_u \sim 10^5$. The SM does not offer an explanation for this hierarchy in the Yukawa couplings.

Several theories beyond the Standard Model (BSM) attempt to address this problem. Among them, multi-Higgs-doublet models ~\cite{Branco:2011iw,Keus:2013hya,Bento:2018fmy,Camargo-Molina:2017klw,Arroyo-Urena:2019lzv,Campos:2017dgc, Blechman:2010cs} allow a lesser degree of hierarchy in the Yukawa couplings compared to the SM. In these models, the Higgs VEV is no longer universal, as more than one neutral Higgs field acquires a non-zero VEV during EWSB. These models present a paradigm where the hierarchy is distributed among the multiple Higgs VEVs as well as the different Yukawa couplings.

In this article, we consider a model where the mass of each SM fermion is induced by the VEV of a unique {\it Private} Higgs (PH) field \cite{Porto:2007ed}, while the Yukawa couplings are naturally {\it democratic}, meaning, all $\mathcal{O}(1)$ numbers. However, the PH model by construction is flavor diagonal and has no explanation for the observed flavor structure of the SM. Hence, we introduce discrete global symmetries and a spectrum of singlet scalar fields and heavy singlet vector-like quarks (VLQs) which act as messengers between the different quark generations and recreate the observed flavor structure in the quark sector of the SM, subject to the imposed symmetries. Interestingly{\cre,} we find that given our model assumptions~($\mathcal{O}$(1) couplings and $m_q\sim v_q$), the induced CKM matrix is \emph{independent} of the SM fermion masses and instead dictated by ratios of our singlet VEVs and the VLQ masses. As is the case with most multi-Higgs models, one has to be careful not to generate large Flavor Changing Neutral Currents (FCNCs). FCNCs, which, for example, induce $D^{0}-\bar{D^{0}}$ mixing, are absent at tree level in the SM and are highly constrained experimentally. We undertake a careful study of possible constraints on our model. While the effects of our model are currently inaccessible, we provide targets for future electroweak precision experiments. Finally, prospects for direct searches at the LHC for our VLQs are discussed.

\section{The Model}\label{sec:model}
We introduce a model that removes the SM hierarchy of dimensionless Yukawa couplings 
and instead generates the observed hierarchy of quark masses and flavor structure through {\it dimensionful}  quantities. The hierarchical VEVs of the PH fields are associated with a hierarchy in the respective (physical) Higgs masses~$M_{H^q}$, and can be thought of as a scalar see-saw mechanism~\cite{Porto:2007ed}, 
\begin{equation}\label{eqMh}
 M_{H^q} \sim v_t \sqrt{\frac{v_t}{v_q}}   \;,
\end{equation}
where $q$ is the SM quark which acquires mass due to the VEV of that particular Higgs field, and $v_t\sim 246$ GeV, the VEV associated with the ``top Higgs."  We assume this ``top Higgs'' to be the observed 125 GeV Higgs boson. Hence, in addition to the SM quarks, $\{Q_{L}^{'j},u_{R}^{j},d_{R}^{j}\}${\cre ,} where $j=\{1,2,3\}$ represents the quark generation, there is a spectrum of six fields. Each of these $SU(2)_{L}$ doublet PH fields, $\{H^{j}_{u},H^{j}_{d}\}$, has $\mathcal{O}(1)$ Yukawa coupling to its corresponding quark.  

The PH model, by its construction, gives rise to purely diagonal couplings of the PH fields to the quarks. Therefore, we introduce a spectrum of singlet-scalar fields, $S$, and heavy singlet vector-like fermions, $\psi$, to obtain the observed masses and flavor structure of the SM. We impose  the following discrete global symmetries which allow only the interactions we want.
\begin{itemize}
\setlength\itemsep{-.2em}
    \item $\mathbb{Z}_{2}^{Q_{i}}$ for SM quarks $\{Q_{L}^{'i},u_{R}^{i},d_{R}^{i}\}$;
    \item $\mathbb{Z}_{2}^{H_{u^i}}$ or $\mathbb{Z}_{2}^{H_{d^i}}$ for the appropriate PH fields~\footnote{Note that this implies the presence of additional singlet scalar fields charged under $\mathbb{Z}_{2}^{H_{d/u^i}}$ which lead to the appropriate pattern of VEV generation from the Higgs potential. Since these fields could potentially contribute to the diagonal couplings of the quarks~(suppressed by the heavy VLQ masses), we further assume that the SM quarks, VLQs and the $S^{ij}$ listed in Table~\ref{Tab:charge} are charged under an additional $\mathbb{Z}_{3}$ to forbid such interactions. We will not  consider these further in this article.} and right-handed quarks;
    \item VLQ quarks $\psi^{ij}$ are charged under $\mathbb{Z}_{2}^{Q_{i}}$ and $\mathbb{Z}_{2}^{H_{u^j}}$, with $i\neq j$~($\psi^{ij}\neq \psi^{ji}$);
    \item  Singlets $S^{ij}$ are charged under both $\mathbb{Z}_{2}^{Q_{i}}$ and $\mathbb{Z}_{2}^{Q_{j}}$, with $i\neq j$~($S^{ij} = S^{ji}$). 
\end{itemize}
 In Table~\ref{Tab:charge}, we summarize the charge assignments of the SM quarks, PH fields, $S$, and $\psi$, under the SM gauge as well as the newly introduced global symmetries.
\begin{table}[!htbp] 
   \begin{center}
\begin{tabular}{|c|c|c|c|c|c|c|c|c|c|}
\hline
&&&&&&&\\
   {}  & ${\bar Q}^{'i}_L$ & $H^j_u$ & $H_d^j$ & ${\psi}^{ij}$  & $S^{ij}$ & $d_R^{i}$&$u_R^{i}$  \\
&&&& $(i\neq j)$ & $(i\neq j)$& &\\   
   \hline
  \hline
   $SU(3)_C$& $\overline{3}$ &$- $&$-$& $3 $ & $-$ & $3$& $3$\\
   \hline
    $SU(2)_L$ & $\overline{2}$ & $2$  & $2$ & $-$ & $-$ & $-$ &$-$ \\
    \hline
    $\mathbb{Z}^{Q_{i}}_{2}$ & $-1$ & $-$  & $-$ & $-1$ & $(-1,-1)$ & $-1$& $-1$\\
    \hline
      $\mathbb{Z}_{2}^{H_{u^j}} $ & $-$ & $-1$ & $-$ &  $-1$ & $-$ & $-$ & $-1^*$\\
      \hline
      $\mathbb{Z}_{2}^{H_{d^j}} $ & $-$ & $-$ & $-1$ &  $-$ & $-$ & $-1^*$ &$-$\\
      \hline
\end{tabular}
\caption{Charge assignments for the different fields under relevant SM and global symmetries: $SU(3)_C$,  $SU(2)_L$,  $\mathbb{Z}_{2}^{Q_{i}}$, $\mathbb{Z}_{2}^{H_{u^j}}$ and $\mathbb{Z}_{2}^{H_{d^j}}$. $^*$ For the SM right-handed quarks $j = i$.  
} \label{Tab:charge}
\end{center}
\end{table}

In the rest of the article, the indices $\{i,j\}$ denote charges under the symmetries listed in Table~\ref{Tab:charge}. We emphasize that the $\psi^{ij}$ are not charged under $\mathbb{Z}_{2}^{H_{d^j}}$, the $\mathbb{Z}_{2}$ associated with the down-type Higgs bosons.  Thus the $\psi^{ij}$ only act as messengers in the up sector. The $\mathbb{Z}_{2}^{Q_{i}}$ symmetry is associated with a generation index $i$  to prevent unwanted interactions between different generations. This symmetry prohibits having any FCNC interactions in the model, both at the tree and one-loop levels~\cite{Chen:2021ftn}, thus decoupling the model from most of the limitations of flavor changing interactions ~\cite{Silvestrini:2019sey,Agrawal:2014aoa,Gherardi:2021pwm}. However, as we discuss later, FCNCs are still induced in the low-energy effective theory due to field redefinitions.

Given these charge assignments, the allowed dimension-4 Lagrangian for our fermion-scalar interactions is given by,
\begin{equation}
\begin{split}
\mathcal{L} &\supset \sum_{i=1,2,3}\left[\left(\lambda_{ii}^{u}{\bar Q}^{'i}_L{\tilde{H}_{u}}^{i}u^i_{R} + \lambda_{ii}^{d} {\bar Q}^{'i}_L H_d^i d_R^i\right) \right.\\
& \qquad\quad 
+ \sum^{i \neq j}_{j=1,2,3}\left(\lambda_{ij}^{u} {\bar Q}^{'i}_L H_u^j\psi_R^{ij} + \alpha_{ij} {\bar\psi}_L^{ij} u_R^j S^{ij}\right. 
+ \left.\left. M_{ij}{\bar \psi}^{ij}_L\psi^{ij}_R \right)\right]
               + \mathrm{h.c.}
\end{split}
\end{equation}
Here, $\tilde{H}=i\sigma_{2}H^{*}$, ${Q}_L^{'}$ are the $SU(2)_L$ weak eigenstates, $\lambda_{ij}$ and $\alpha_{ij}$ are arbitrary, complex, coupling constants, and  $M_{ij}$ are the masses of the VLQs. Since right-handed quarks do not interact weakly, we assume each PH doublet couples a left-handed interaction quark to a right-handed physical (mass) quark ($u_R^i$ or $d_R^i$). From the above, we see that the singlet-scalar fields can interact with both the $\psi$ and the right-handed SM quarks. The $\psi$ on the other hand, can interact additionally with the PH bosons and the SM quark doublets.

Since the VLQs in our model are expected to be heavy, we integrate them out to obtain the low-energy effective Lagrangian which we will match onto the SM Lagrangian. Including terms up to dimension-5, we obtain, 
\begin{flalign}
\mathcal{L} & =\sum_{i=1,2,3} \left[\left( \lambda_{ii}^{u} {\bar Q}^{'i}_L \tilde{H}_u^i u_R^i + \lambda_{ii}^{d}{\bar Q}^{'i}_L H^i_d d^i_R\right) 
+\sum^{j \neq i}_{j=1,2,3} \left(\frac{-\lambda_{ij}^u}{M_{ij}}\alpha_{ij} {\bar Q}^{'i}_L H_u^j u^j_{R} S^{ij}\right)\right]+ \mathrm{h.c.}\;.
\label{eq5}
\end{flalign}

\begin{figure}[tb!]
\begin{center}
    \centering
    \includegraphics[scale=0.5]{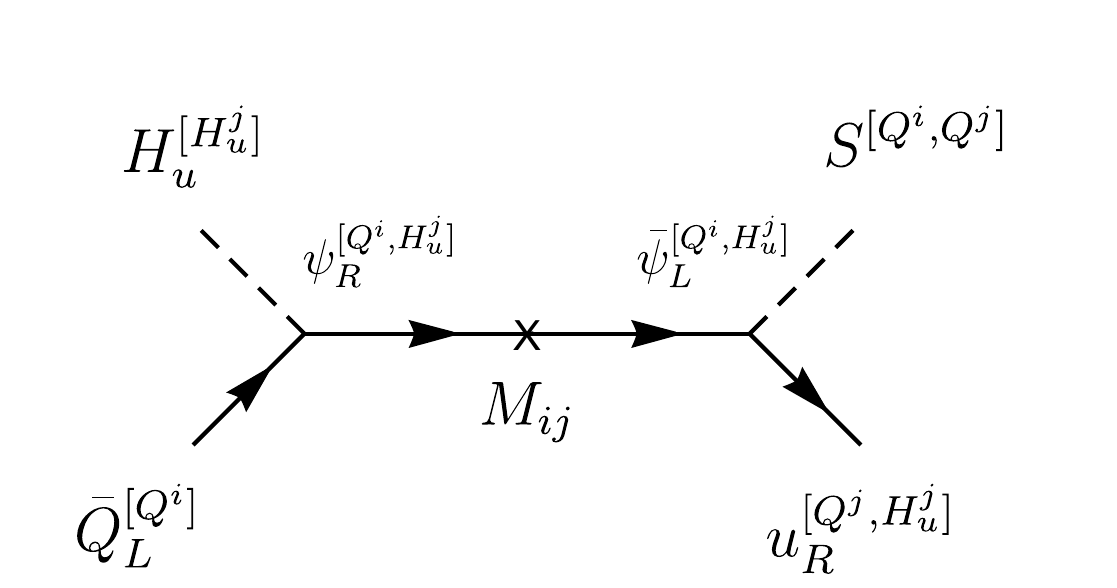}
     \caption{An example Feynman diagram for the VLQs and singlets acting as messengers between different generations in order to generate the SM quark mass matrix after EWSB. Here $H_{u}$ indicates a PH field, $Q'_L$ represents a left-handed SM quark, $\psi$ represents a VLQ, $S$ represents a singlet scalar, and $u_{R}$ represents a right-handed up-type SM quark. $M_{ij}$ is the mass of the VLQ. The superscripts of each field represent its charge assignments under all the $\mathbb{Z}_{2}$ symmetries listed in Table~\ref{Tab:charge}.}
    \label{fig:8}
\end{center}
\end{figure}

In Fig.\ref{fig:8} we show an example Feynman diagram where integrating out the VLQs generates the mass matrix for the SM quarks after EWSB. For clarity, we have explicitly written the $\mathbb{Z}_{2}$ symmetry as the superscript label that each particle is charged under. While these interactions are sufficient to generate the observed CKM matrix, they do not provide us with any further information to probe the high{\cre -}energy structure of our model. 

The next higher-order corrections come from the kinetic terms for the VLQs,  
\begin{flalign}
\bar{\psi}^{ij}\slashed{D}_R\psi^{ij} \rightarrow \frac{|\lambda^{u}_{ij}|^{2}}{M_{ij}^{2}}\left({\bar Q}^{'i}_L \tilde{H}_u^j\right)\slashed{D}_R\left(Q^{'i}_L H_u^j\right) 
+~\frac{|\alpha_{ij}|^{2}}{M_{ij}^2}{\bar u}_R^j S^{*ij} \slashed{D}_R u_R^j S^{ij}\;.
\label{eq6}
\end{flalign}
These corrections generate the dimension-6 operators in the Lagrangian. Here the subscript $R$ on $\slashed{D}_R$ denotes that the covariant derivative is the same as that for the SM right-handed up-type quarks which have the same SM quantum numbers as our VLQs. 

After EWSB, the relevant Lagrangian up to dimension-6 (combining Eqs.~\ref{eq5} and \ref{eq6}), is given by 
\begin{eqnarray}
\mathcal{L} &=& \sum_{i=1,2,3} \Bigg[\left( \lambda_{ii}^{u} v^i_u {\bar Q}^{'i}_L u_R^i + \lambda_{ii}^d v^i_d {\bar Q}^{'i}_L d_R^i\right) 
+~ \sum^{i \neq j}_{j=1,2,3} \left(\frac{-\lambda_{ij}^{u}\alpha_{ij}}{M_{ij}} v_u^j s^{ij} {\bar Q}^{'i}_L u_R^j \right. \nonumber\\
&& \qquad \qquad  
+~ \left.\frac{|\lambda^{u}_{ij}|^{2}}{M_{ij}^{2}}|{v_{u}^{j}}|^{2} 
{\bar Q}^{'i}_L \slashed{D}_R Q^{'i}_L \right. 
+~ \left.\left.\frac{|\alpha_{ij}|^{2}}{M_{ij}^{2}}|{s^{ij}}|^{2} {\bar u}_R^j \slashed{D}_R u_R^j\right) \right] +  {\rm KE_{SM}}\; + \mathrm{h.c.} ,
\label{eq9}
\end{eqnarray}
where $v_{u/d}^i$ and $s^{ij}$ are the VEVs of the PH fields and singlets, and  ${\rm KE_{SM}}$ denotes the SM kinetic terms.

From the above, we see that the derivative terms give non-standard contributions to the kinetic terms for the quarks. Putting them all together, for each generation $i$, and only considering the up sector,
\begin{eqnarray}
\mathcal{L}^i_{\rm{KE}} &=& {\sum^{j\neq i}_{j=1,2,3}}\left(\frac{|\lambda^{u}_{ij}|^{2}}{M_{ij}^{2}}{|v_u^j|}^{2}{\bar Q}^{'i}_L \slashed{D}_R Q^{'i}_L \right) + {\bar Q}^{'i}_L \slashed{D}_L Q^{'i}_L \nonumber \\
&
+& \sum^{j\neq i}_{j=1,2,3}\left(\frac{|\alpha_{ij}|^{2}}{M_{ij}^{2}}{|s^{ij}|}^2 {\bar u}_R^i \slashed{D}_{R} u_R^i \right) + {\bar u}^i_R \slashed{D}_R u^i_R\,,~~
\end{eqnarray}
where $\slashed{D}_{R}$ and $\slashed{D}_{L}$ are the covariant derivatives for the right- and left-handed up-type quarks respectively,
\begin{equation}
\begin{aligned}
D^{L}_{\mu}&= \partial_{\mu}-ieA_{\mu}Q+ig_3\left(G^{\alpha}_{\mu}t_{\alpha} \right) 
                        -\frac{ig}{\sqrt{2}}(W^{+}_{\mu}T^{+}+W^{-}_{\mu}T^{-})
                         -\frac{ig}{\cos{\theta_{w}}}Z_{\mu}(T^{3}-\sin^{2}\theta_{w}Q)\;,
\end{aligned}
\end{equation}

\begin{equation}
\begin{aligned}
D^{R}_{\mu}&= \partial_{\mu}-ieA_{\mu}Q+ig_3\left(G^{\alpha}_{\mu}t_{\alpha} \right)
                       +\frac{ig}{\cos{\theta_{w}}}Z_{\mu}\sin^{2}\theta_{w}Q\;.
\end{aligned}
\end{equation}
Explicitly expanding the derivative terms, we find,
\begin{equation}
\begin{aligned}
\mathcal{L}^i_{\rm{KE}}&=\left(1+{\sum^{i \neq j}_{j=1,2,3}}\frac{|\lambda^{u}_{ij}|^{2}}{M_{ij}^{2}}{|v_{u}^{j}|}^{2}\right) \left[\partial_{\mu}-ieA_{\mu}Q +ig_3\left(G^{\alpha}_{\mu}t_{\alpha} \right)\right] 
~{\bar Q}^{'i}_L \gamma^\mu Q^{'i}_L \\
&-\frac{ig}{\cos{\theta_{w}}}\left[ \left(T^{3}-\sin^{2}\theta_{w}Q\right)-\sin^2\theta_w{\sum^{i \neq j}_{j=1,2,3}}\frac{|\lambda^{u}_{ij}|^{2}}{M_{ij}^{2}}{|v_{u}^{j}|}^{2}\right] 
~Z_{\mu}{\bar Q}^{'i}_L \gamma^\mu Q^{'i}_L \\
& -\frac{ig}{\sqrt{2}} \left(W^{+}_{\mu}T^{+}+W^{-}_{\mu}T^{-}\right){\bar Q}^{'i}_L \gamma^\mu Q^{'i}_L \\
&+\left(1+\sum^{j\neq i}_{j=1,2,3}\frac{|\alpha_{ij}|^{2}}{M_{ij}^{2}}{|s^{ij}|}^{2}\right)\left[\partial_{\mu}-ieA_{\mu}Q+ig_3\left(G^{\alpha}_{\mu}t_{\alpha} \right)
+\frac{ig}{\cos{\theta_{w}}}Z_{\mu}\sin^{2}\theta_{w}Q\right]\bar{u}_{R}^{i}\gamma^\mu{u}_{R}^{i}\;.
\end{aligned}
\label{eq13}
\end{equation}
By inspection of Eq.~\ref{eq13}, we see that the photon, the gluon, and the $Z$ couplings of the right-handed quarks scale similarly to their respective derivative terms. Thus, these couplings do not get corrections when canonically normalized. However, both the $W$ and $Z$ couplings of the left-handed quarks do get corrected. For the $W$ couplings, the rescalings simply imply a redefinition of the CKM matrix. However, the non-universal rescalings for the $Z$ couplings lead to non-trivial FCNCs and may allow our model to be probed at near-future experiments.

After EWSB, the mass term for the up-type quarks in the interaction basis is given by,
\begin{equation}
    \mathcal{L}_{\rm mass} = \bar{u}_L' \Lambda_u V_u u_R\;, 
\end{equation}
where the elements of $\Lambda_u$ and $V_u$ are defined according to Eq.~\ref{eq9}: $\Lambda_u^{ii}=\lambda_{ii}^u$,  $\Lambda_u^{ij} = \lambda_{ij}^u \alpha_{ij} s^{ij}/M_{ij}$, and $V_{u}^{ij} = \delta^{ij} v_u^i$ is a diagonal matrix of the PH VEVs.  

We canonically renormalize the Lagrangian by redefining the ${u'_{L}}^{i}$ and ${u_{R}}^{i}$ fields as follows,
\begin{eqnarray}\label{QL1}
{u''_{L}}^{i} &=&  {u'_{L}}^{i}\sqrt{1+\sum^{j \neq i}_{j=1,2,3}\frac{|\lambda^{u}_{ij}|^{2}}{M_{ij}^{2}}{|v_{u}^{j}|}^{2}}\,, \\
\label{uR1}
{u''_{R}}^{i} &=&{u_{R}}^{i}\sqrt{1+\sum^{j \neq i}_{j=1,2,3}\frac{|\alpha_{ij}|^{2}}{M_{ij}^{2}}{|s^{ij}|}^{2}}\,.
\end{eqnarray}
The above redefinitions can also be written in matrix form as,
\begin{equation}\label{eqAB}
u''_{L}=A~{u'_{L}}\;,  \qquad\qquad u''_{R}=B~{u}_{R}\;,
\end{equation}
where ${u''_{L}}^{i}$ and ${u''_{R}}^{i}$ are the new fields, while $A$ and $B$ are diagonal matrices defined according to the rescalings given in Eqs.~\ref{QL1} and \ref{uR1}. Hence the canonically-normalized mass term for the interaction basis fields is
\begin{equation}
\mathcal{L}_{mass} ~=~{\bar{{u}}''_{L}}~A^{-1}~\Lambda_u ~V_u ~B^{-1} u''_{R}\;.
              \label{eqmass}
\end{equation}
Generally, the diagonal physical mass matrix is obtained by bi-unitary rotations of the interaction mass matrix. However, because the right-handed quarks do not interact weakly, we assume the right-handed rotation matrices to be the identity matrix. Hence the diagonal physical mass matrix is given by
\begin{equation}
   M_{u}=U^{\dagger}_{L}A^{-1}\Lambda_u V_uB^{-1}\;,
    \label{eq:15}
\end{equation}
and therefore,
\begin{equation}
    U^{\dagger}_{L}=M_{u}BV_{u}^{-1}{\Lambda_u}^{-1}A\;.
    \label{eq:15}
\end{equation}
Since the down sector is diagonal, we find that the CKM matrix is given by,
\be
V_{\rm CKM} ~= ~U^{\dagger}_{L}A^{-1} ~= ~M_{u}BV_{u}^{-1}{\Lambda_u}^{-1}.
\label{eq29}
\ee
It is worth noting that $M_{u},~ B$, and $V^{-1}_{u}$ are all diagonal matrices. Since $m_q\sim v_q$ 
the product of $M_{u}$ and $V^{-1}_{u}$ may be assumed to be 
$\sim$ the identity matrix. Additionally, $B$ is a small redefinition that arises from the right-handed quarks, so we can expect  $M_{u}BV^{-1}_{u}$ also to be very close to the identity matrix. As a result, we can approximate $V_{\rm CKM}\approx{\Lambda}_{u}^{-1}$.

\section{The Spectrum}\label{sec:spect}
We know experimentally that $V_{\rm CKM}$ is {\it almost} but not quite diagonal, and the small off-diagonal elements encode the misalignment between the rotations of the up and down sectors. Some of the questions that arise are, if these two rotations are in independent sectors, why are they so similar? If they are related, why is the difference so small? Further, a common assumption is that the third generation is almost decoupled in $V_{\rm CKM}$ due to the SM fermion mass hierarchies. 

Our model is constructed to precisely address these concerns. To understand the hierarchical CKM structure generated in our model, We first look at the relationship between the VLQ masses and the other parameters in the model. We know from Eq.(\ref{eq29}),
\begin{equation}
V^{-1}_{\rm CKM} = {\Lambda_u}V_{u}B^{-1}{M_{u}}^{-1}. 
\label{eqckm}
\end{equation}
The right-hand side of this can be expressed in terms of our model parameters as,
\begin{equation}
\begin{bmatrix}\displaystyle \frac{\lambda_{11}v_{u}}{xm_{u}}  & \displaystyle  \frac{\lambda_{12}\alpha_{12}v_{c}s_{12}}{ym_{c}M_{12}}  & \displaystyle \frac{\lambda_{13}\alpha_{13}v_{t}s_{13}}{zm_{t}M_{13}}\\ 
\\
\displaystyle \frac{\lambda_{21}\alpha_{21}v_{u}s_{21}}{xm_{u}M_{21}}& \displaystyle \frac{\lambda_{22}v_{c}}{ym_{c}}& \displaystyle\frac{\lambda_{23}\alpha_{23}v_{t}s_{23}}{zm_{t}M_{23}}\\
\\
\displaystyle \frac{\lambda_{31}\alpha_{31}v_{u}s_{31}}{xm_{u}M_{31}}& \displaystyle \frac{\lambda_{32}\alpha_{32}v_{c}s_{32}}{ym_{c}M_{32}}& \displaystyle \frac{\lambda_{33}v_{t}}{zm_{t}}
\label{eqmat}
\end{bmatrix}\;,
\end{equation}\\
where $x,y$ and $z$ are defined as,
\begin{equation}
	\begin{aligned}
    x=\sqrt{1+\left|\frac{\alpha_{12}s_{12}}{M_{12}}\right|^{2}+\left|\frac{\alpha_{13}s_{13}}{M_{13}}\right|^{2}}\;,\\ 
		y=\sqrt{1+\left|\frac{\alpha_{21}s_{21}}{M_{21}}\right|^{2}+\left|\frac{\alpha_{23}s_{23}}{M_{23}}\right|^{2}}\;,\\
		z=\sqrt{1+\left|\frac{\alpha_{31}s_{31}}{M_{31}}\right|^{2}+\left|\frac{\alpha_{32}s_{32}}{M_{32}}\right|^{2}}\;.\\
	\end{aligned}
	\label{eqxyz}
\end{equation}

While we have many degrees of freedom, for simplicity, we take Yukawa couplings $\lambda_{ij}\simeq\lambda_{ii}$, singlet VEVs $s_{ij} = s_{ji}${\cre ,} and VLQ masses $M_{ij} \neq M_{ji} $. Further we assume all $|\alpha_{ij}|$  and $|\lambda_{ij}|$ couplings to be $\mathcal{O}{(1)}$ and $|v^i_u| \approx m^i_u$. Under these assumptions, and ignoring possible phases\footnote{The coupling constants, $\lambda^u_{ij}$, $\alpha_{ij}$, and the PH and singlet VEVs, $v_u^j$, $s^{ij}$, are all potentially complex quantities which would combine to produce the required CP violating phase in $V_{\mathrm{CKM}}$. However, this induced CP-violating phase does not play a role in the hierarchy generated in the quark masses and mixing angles in our model. Therefore, in the remainder of this paper we remain agnostic about possible phases in these quantities. Further, it should be understood that we are explicitly only considering their modulus when referring to these quantities below.},
\begin{equation}
	|V^{-1}_{\rm CKM}|\simeq	\begin{bmatrix}\displaystyle \frac{1}{x}&\displaystyle\frac{s_{12}}{M_{12}y}&\displaystyle\frac{s_{13}}{M_{13}z}\\
    \\
		\displaystyle\frac{s_{12}}{M_{21}x}&\displaystyle\frac{1}{y}&\displaystyle\frac{s_{23}}{M_{23}z}\\
        \\
		\displaystyle\frac{s_{13}}{M_{31}x}&\displaystyle\frac{s_{23}}{M_{32}y}&\displaystyle\frac{1}{z}
	\end{bmatrix}\;,
	\label{eq:CKMinvmod}
\end{equation}
and 
\begin{equation}
	\begin{aligned}
		x \approx \frac{1}{|V^{-1}_{\rm CKM}|_{11}}=\frac{1}{\mathcal{V}_{11}},\\ 
		y \approx \frac{1}{|V^{-1}_{\rm CKM}|_{22}}=\frac{1}{\mathcal{V}_{22}},\\
		z \approx \frac{1}{|V^{-1}_{\rm CKM}|_{33}}=\frac{1}{\mathcal{V}_{33}}.\\
	\end{aligned}
	\label{eqxyzmod}
\end{equation}
Combining Eqs.~\ref{eq:CKMinvmod} and \ref{eqxyzmod}, we find
\begin{equation}
	|V^{-1}_{\rm CKM}|\simeq	\begin{bmatrix}\displaystyle \mathcal{V}_{11}& \displaystyle \mathcal{V}_{22}\frac{s_{12}}{M_{12}}& \displaystyle \mathcal{V}_{33}\frac{s_{13}}{M_{13}}\\
    \\
		\displaystyle\mathcal{V}_{11}\frac{s_{12}}{M_{21}}& \displaystyle\mathcal{V}_{22}&\displaystyle\mathcal{V}_{33}\frac{s_{23}}{M_{23}}\\
	\\
       \displaystyle \mathcal{V}_{11}\frac{s_{13}}{M_{31}}&\displaystyle\mathcal{V}_{22}\frac{s_{23}}{M_{32}}&\displaystyle\mathcal{V}_{33}
	\end{bmatrix}\;.
	\label{eq4.34}
\end{equation}
 We see from Eqs.~\ref{eqmat}~-~\ref{eq4.34} that the rotation matrix, and hence also the CKM matrix, is not only completely dictated by the structure of our model, but also \emph{independent} of the fermion masses. The latter follows from the imposition of $v^i_u \approx m^i_u$. Further, we obtain very simple relationships between the CKM elements, the VLQ masses, and the singlet VEVs,
\begin{align}
	&M_{12}\approx s_{12}\frac{\mathcal{V}_{22}}{\mathcal{V}_{12}}, \quad M_{21} \approx s_{21}\frac{\mathcal{V}_{11}}{\mathcal{V}_{21}},\nonumber\\
	&M_{23}\approx s_{23}\frac{\mathcal{V}_{33}}{\mathcal{V}_{23}}, \quad M_{32}\approx s_{23}\frac{\mathcal{V}_{22}}{\mathcal{V}_{32}},\nonumber\\
	&M_{31}\approx s_{31}\frac{\mathcal{V}_{11}}{\mathcal{V}_{31}}, \quad M_{13}\approx s_{13}\frac{\mathcal{V}_{33}}{\mathcal{V}_{13}}.
	\label{eq4.37}
\end{align}

The experimentally measured values of $V_{\rm CKM}$ are~\cite{ParticleDataGroup:2024cfk},
\begin{equation}
	|V_{\rm CKM}|=\begin{bmatrix}
		0.97435 &\qquad 0.22501  &\qquad 0.003732  \\
		0.22487  &\qquad 0.97349 &\qquad 0.04183  \\
		0.00858 &\qquad 0.04111   &\qquad 0.999118 \\
	\end{bmatrix}\;,
	\label{eqckml}
\end{equation}
which lead to the following relationships,
\begin{eqnarray}
 M_{12} &\approx& M_{21} ~\approx~ 4\,s_{12}\,,\nonumber\\
 M_{23} &\approx& M_{32} ~\approx~ 25\,s_{23}\,,\nonumber\\
 M_{31} &\approx& 1000\,s_{31}\,,\,M_{13} ~\approx~ 150\,s_{13}\,.
\label{eqMS}
\end{eqnarray}

Returning now to the $Z$ couplings, we demonstrate the relationships between the singlet VEVs and the FCNCs induced by the field redefinitions in Eq.~\ref{eqAB}. Focusing on the relevant left-handed fields, we find,
\begin{eqnarray}
\mathcal{L}_{Z} &=& Z_{\mu}\bar{u}_{L}\gamma^\mu U_L^\dagger A^{-1} CA^{-1}U_L{u_{L}}\;, \nonumber \\
&=& Z_{\mu}{\bar u}_{L}\gamma^\mu ~V_{\rm CKM}~C~V_{\rm CKM}^\dagger ~{u_{L}}\;, 
\label{eqz}                      
\end{eqnarray}
where the unprimed fields are in the mass basis and $C$ is a diagonal matrix of non-universal couplings. Proceeding from Eq.~\ref{eq13}, the matrix $C$ can be expressed as follows,
\begin{equation}
C_{ij}=\delta_{ij} ~g^{SM}_{Z\bar{q}_Lq_L}
                    \left(1-\frac{ \sin^{2}\theta_{w}Q{\sum^{i \neq j}_{j=1,2,3}}\frac{|\lambda^{u}_{ij}|^{2}}{M_{ij}^{2}}{|v^{u}_{j}|}^{2}}{T^{3}-\sin^{2}\theta_{w}Q}\right)\;,
                    \label{eqgg}
\end{equation}
where $g^{SM}_{Z\bar{q}_Lq_L}=\frac{-ig}{\cos{\theta_{w}}}(T^{3}-\sin^{2}\theta_{w}Q)$. For reference, $g^{SM}_{Z\bar{q}_Rq_R}=\frac{ig}{\cos{\theta_{w}}}(\sin^{2}\theta_{w}Q)$. A priori, the elements of $C$ depend on the VLQ masses and the PH VEVs~(quark masses). However, as shown in Eq.~\ref{eqMS}, assuming $\mathcal{O}(1)$ couplings, the CKM matrix imposes relationships between the VLQ masses and the singlet VEVs.  Hence, $C$ can be written just in terms of the singlet VEVs and SM parameters as,
 \begin{align}
 	C_{11}&\approx g^{SM}_{Z\bar{q}_Lq_L}\left[1- \frac{\sin^{2}{\theta_{w}}Q}{T^{3}-\sin^{2}{\theta_{w}}}\left(\left|\frac{m_{t}\mathcal{V}_{13}}{s_{13}\mathcal{V}_{33}}\right|^{2}+\left|\frac{m_{c}\mathcal{V}_{12}}{s_{12}\mathcal{V}_{22}}\right|^{2}\right)  \right],\\
 	C_{22}&\approx g^{SM}_{Z\bar{q}_Lq_L}\left[1- \frac{\sin^{2}{\theta_{w}}Q}{T^{3}-\sin^{2}{\theta_{w}}}\left(\left|\frac{m_{u}\mathcal{V}_{21}}{s_{21}\mathcal{V}_{11}}\right|^{2}+\left|\frac{m_{t}\mathcal{V}_{23}}{s_{23}\mathcal{V}_{33}}\right|^{2}\right)  \right],\\
 	C_{33}&\approx g^{SM}_{Z\bar{q}_Lq_L}\left[1- \frac{\sin^{2}{\theta_{w}}Q}{T^{3}-\sin^{2}{\theta_{w}}}\left(\left|\frac{m_{u}\mathcal{V}_{31}}{s_{31}\mathcal{V}_{11}}\right|^{2}+\left|\frac{m_{c}\mathcal{V}_{32}}{s_{32}\mathcal{V}_{22}}\right|^{2}\right)  \right].
 \end{align}
Using the known numerical values for the CKM elements, the weak-mixing angle ($\sin^2\theta_w=0.2313$~\cite{ParticleDataGroup:2024cfk}), the weak coupling constant ($g=e/\sin\theta_w\simeq 0.635$~\cite{ParticleDataGroup:2024cfk}), the third component of isospin for the up-type quarks, and the quark masses, we find that the deviation from the SM (including FCNC) to the $Z$ coupling matrix, $V_{\rm CKM} C V_{\rm CKM}^\dagger$, is given by, 
\begin{equation}\label{eqZqq}
	g^u_{Z\bar q_Lq_L} ~=~ g^{\rm SM}_{Z\bar{q}_Lq_L}~\mathbb{I} ~+~\begin{blockarray}{c  c c c}
		& u_L & c_L & t_L\\
		\begin{block}{c@{\hspace{9pt}} [c c c]}
			\bar{u}_L &\left(\frac{0.1 }{s_{13}^{2}}\text{GeV}^{2}+\frac{0.32}{s_{23}^{2}}\text{GeV}^{2}\right) & \frac{1.37}{s_{23}^{2}}\text{GeV}^{2}& ~-\frac{0.06}{s_{23}^{2}}\text{GeV}^{2}~ \\
            \\
			\bar{c}_L &\frac{1.37}{s_{23}^{2}}\text{GeV}^{2} & \frac{6}{s_{23}^{2}}\text{GeV}^{2} & -\frac{0.25}{s_{23}^{2}}\text{GeV}^{2}\\
            \\
			\bar{t}_L &-\frac{0.06}{s_{23}^{2}}\text{GeV}^{2} &-\frac{0.25}{s_{23}^{2}}\text{GeV}^{2} &  \frac{0.01}{s_{23}^{2}}\text{GeV}^{2}~\\
		\end{block}
	\end{blockarray}\;\;,
\end{equation}
where $g^{\rm SM}_{Z\bar{q}_Lq_L} = -0.2367$ and $\mathbb{I}$ is the identity matrix. We have labeled the rows and columns with the relevant quarks and dropped subdominant terms for clarity.

\section{Experimental Constraints}\label{sec:exp}
\subsection{Higgs Phenomenology}
In our model, $v_q\sim m_q \ll v_t \sim m_t$. As a consequence of this hierarchy and Eq.~\ref{eqMh}, the corresponding Higgs masses are hierarchically heavier than the top Higgs ($H_t=H_{125}$). The lightest of the remaining PHs is the ``bottom" Higgs with mass $M_{H^b}\sim 1.5$ TeV. In this setup, thus, $H_t$ essentially decouples from the other PHs. It has very SM-like couplings, consistent with current experimental observations. However, small deviations from SM-like couplings arise at higher order, $\sim v_t^2/M_{H^j}^2 \sim v_j/v_t$ and can be of either sign~\cite{BENTOV_2013}. For example, the largest coupling deviations expected would be to $H_tb\bar{b} \sim \mathcal{O}$ (few \%).

We note here that FCNC interactions, proportional to possible confirmation of non-untarity~\cite{Belfatto:2019swo, ParticleDataGroup:2024cfk} in the CKM matrix, are also induced in the Higgs couplings at dimension-7. This is unlike what happens in generic VLQ models where the $Z$ and $H$ FCNCs occur at the same order~\cite{Alves_2024}. Instead, it is due to the fact that our CKM is generated at dimension-5 rather than dimension-4. Further, since the dimension-7 operators are cubic in the Higgs fields,  the $H_qq\bar{q}$ couplings can be expected to deviate from their tree-level values. However, such deviations would lead to only  sub-percent level deviations in the $H_tb\bar{b}$ coupling. Current measurements are only sensitive to $\sim$ 20\% coupling deviations in $H _tb\bar{b}$~\cite{CMS:2022dwd, ATLAS:2022vkf} and the HL-LHC is expected to be able to probe at the level of $\sim$ 5\%. Future colliders, such as the ILC or the muon collider, are projected to be able to probe the $H_tb\bar{b}$ coupling at the percent level~\cite{Dawson:2022zbb} and hence may be able to probe our model. 

There is a strong program of searches at the LHC looking for heavy Higgs bosons that arise in many BSM models such as 2 Higgs Doublet models or Supersymmetry. While the experimental limits are sensitive to model parameters and final states, the strongest constraints are obtained for a heavy Higgs produced via gluon fusion and decaying to a pair of $\tau$ leptons. Current $\tau\tau$ channel measurements are able to probe a heavy Higgs of the order of $M_H\sim 2$~TeV~\cite{ATL-PHYS-PUB-2024-008}. While the $H_b$ mass is within reach of the $\tau\tau$ limits, we expect $H_b$ to decay dominantly to a pair of $b$ quarks, for which the search limits are at less than a TeV. $H_b$ can also decay to a pair of $\tau$ leptons via mixing with $H_\tau$, but this is expected to be small $\sim v_bv_\tau/[v_t(v_b-v_\tau)]\sim 0.02$. Hence the $\tau\tau$ search limits are significantly degraded and do not currently provide a constraint on $M_{H_b}$. However, HL-LHC is expected to be able to probe heavy Higgs bosons as heavy as 2.5 TeV~\cite{Dainese:2019rgk}, and hence may be able to probe the possible presence of $H_b$.     

Finally, if the masses of the singlets are of the same order as the SM Higgs, they may potentially provide a means for investigating this model directly at the LHC in the near future. However, the specifics of this scenario depend heavily on the characteristics of the Higgs sector. The Higgs sector of our model is mostly unconstrained at this time and further investigation is beyond the scope of this article.

\subsection{$Z$ Decays, Top Decays, and FCNCs}
The $Z$ decay width as well as other measurements such as deep inelastic scattering cross-sections and forward-backward asymmetries \cite{ParticleDataGroup:2024cfk},  constrain the $Z q\bar{q}$ vector and axial couplings, $g^q_V = (g^{\rm SM}_{Z\bar{q}_Lq_L}+g^{\rm SM}_{Z\bar{q}_Rq_R})/2 = (T^3-2 Q\sin^2\theta_w)$ and $g^q_A =(g^{\rm SM}_{Z\bar{q}_Lq_L}  - g^{\rm SM}_{Z\bar{q}_Rq_R})/2 = T^3$ respectively. In our model, we expect the singlet VEVs to be $\mathcal{O}(100\text{~GeV})$. When $s_{13}\sim s_{23}\gtrsim 100\text{~GeV}$, from Eq.~\ref{eqZqq} we see that the corrections to the $Z\bar u_Lu_L$ coupling is $\lesssim 10^{-4}$; similarly for $Z\bar c_Lc_L$. Hence corrections are negligible and do not contribute to possible deviations in the experimentally measured value of $g^u_V=(0.266 \pm 0.034$)~\cite{ParticleDataGroup:2024cfk}.~\footnote{ Using the best-fit value for $\sin^2\theta_w = 0.23129(4)\pm 0.0017$~\cite{ParticleDataGroup:2024cfk} to compute the SM value of $g^u_V$, one finds a $\sim 2~\sigma$ discrepancy with the best-fit value $g^u_V$ quoted here. However, the corrections induced in our model are too small to be able to account for this discrepancy.} We note that an order of magnitude improvement in the measurement of these couplings may allow us to probe the lowest VEV scale in our model. 

The off-diagonal couplings in Eq.~\ref{eqZqq} are responsible for FCNC interactions involving the $Z$ boson. Hence, in our model, FCNCs are only present in the up-sector and depend only on the singlet VEVs. The strongest constraints on these FCNC couplings come from  $D^0$--${\bar D}^0$ mixing~\cite{Falk:2001hx} and decays.
\begin{figure}[!tbp]
\begin{center}
\includegraphics[width=9cm, height=5cm]{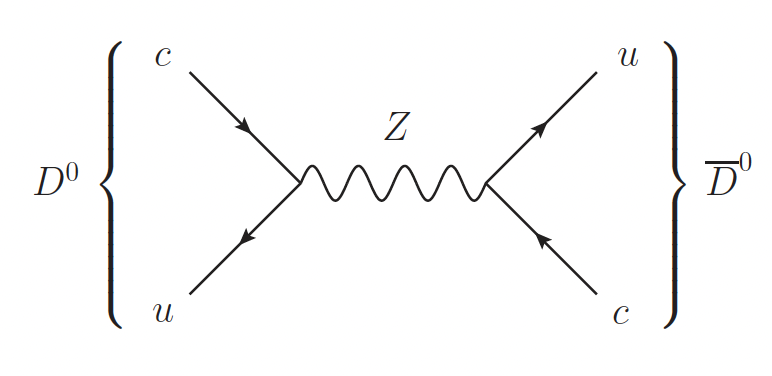}
\caption{$Z$ mediated FCNC $D^{0}-\bar{D}^{0}$ mixing at tree level, taken from Ref.~\cite{Branco:2021vhs}. }
\label{D}
\end{center}
\end{figure}

The tree-level contribution for  $D^0$--${\bar D}^0$ mixing is shown in Fig.~(\ref{D}). The effective Lagrangian is given by \cite{Branco:2021vhs},
\begin{equation}
\mathcal{L}=-\frac{G_{F}}{\sqrt{2}}|g_{Z\bar u_Lc_L}^{u}|^{2}(\bar{u}_{L}\gamma^{\mu}c_{L})(\bar{u}_{L}\gamma_{\mu}c_{L})\;,
\end{equation}
where $G_{F}$ and $g_{Z\bar u_Lc_L}^{u}$ are the Fermi constant and FCNC coupling respectively. This contributes to the $D^0$--${\bar D}^0$ mixing parameter $x_{D}=\Delta m_{D}/\Gamma_{D}$,
given by~\cite{Branco:2021vhs},
\begin{equation}\label{eqxD}
x_{D}\simeq -\frac{\sqrt{2}~m_{D}}{3~ \Gamma_{D}}\,G_{F}\,f^2_D\,B_D\,r(m_c,M_Z)\,|g_{Zu_L\bar{c}_L}^u|^{2}\;,
\end{equation}
where $m_D=(1864.83\pm 0.05)$~MeV, $\Gamma_D=1.6\times10^{-9}$~MeV,  $G_{F}=1.17\times 10^{-5}$ GeV$^{-2}$ \cite{ParticleDataGroup:2024cfk}, and the factors $r(m_{c},M_{Z}) \simeq 0.778$, $B_{D}\simeq 1.18^{+0.07}_{-0.05}$ \cite{Buras:2010nd}, and $f_{D}\simeq 212.0 \text{~MeV}$ \cite{Branco:2021vhs}. 
The experimentally measured value for this parameter, $x_{D}=0.434^{+0.126}_{-0.139}~\%$ \cite{ParticleDataGroup:2024cfk}, constrains the FCNC coupling, $g^{u}_{Z\bar{u}_Lc_L}$, which depends on $s_{23}$ in our model.  Using singlet VEVs $\sim 100$ GeV, then, we find that the contribution to $x_D$ in our model is $\sim 0.1\%$. Substantial hadronic uncertainties make the SM calculation for  observables in $D^0$--${\bar D}^0$ mixing challenging~(see for  example, Ref.~\cite{Dulibic:2024tpo}) and currently large discrepancies exist. However, our model contribution is  negligible and cannot contribute to the resolution of any such discrepancy.

The FCNC coupling, $g^{u}_{Z\bar{u}_Lc_L}$, together with $Z$ decays to leptons can potentially lead to leptonic decays of the $D$ mesons. Such effects are often parameterized in terms of Wilson coefficients. In our model, the relevant effective Lagrangian would be~\cite{Bause:2019vpr, Gisbert:2020vjx},
\begin{equation}
\mathcal{L} = -4\frac{G_F}{\sqrt{2}}\frac{\alpha_e}{4\pi}\left( C_9^\ell O_9^\ell+ C_{10}^\ell O_{10}^\ell   \right)\;,
\end{equation}
    where
    \begin{align}
        O_9^\ell=\left(\bar{u}\gamma_\mu P_L c \right) \left( \bar{l}\gamma^\mu l \right)\;,\qquad \mathrm{and} \qquad 
        O_{10}^\ell=\left(\bar{u}\gamma_\mu P_L c \right) \left( \bar{l}\gamma^\mu \gamma^5l \right)\;.
    \end{align}
    Integrating out the $Z$ and matching to our model Lagrangian parameters from Eq.~\ref{eqZqq},
\begin{align}
C_9^{\ell} &
=
\frac{2\pi}{\alpha_e}\,
g_V^\ell\,
\frac{2\cos \theta_w}{g}\,
g^{u}_{Z\,\bar{u}_L c_L} \simeq -82 \,
g^{u}_{Z\,\bar u_L c_L} \sim -\frac{10^2}{s_{23}^2},
\qquad  \label{eq:C9}\\
C_{10}^{\ell} &
=
\frac{2\pi}{\alpha_e}\,
g_A^\ell\,
\frac{2\cos \theta_w}{g}\,
 g^{u}_{Z\,\bar u_L c_L} \simeq -10^3 \,
g^{u}_{Z\,\bar u_L c_L} \sim -\frac{10^3}{s_{23}^2}, \label{eq:C10}
\end{align}
where the vector and axial couplings for a charged lepton~\footnote{Similar operators also exist for the neutrinos~\cite{Hiller:2025zgr}. However, the experimental constraints are considerably weaker than for the charged leptons~\cite{BESIII:2021slf, Belle:2016qek}.} are given by  $g_V^\ell = -\frac{1}{2} + 2 \sin \theta_w^2=-0.03742$ and 
$g_A^\ell = -\frac{1}{2}$ respectively, and $\alpha_e^{-1}(m_Z)=128.946$. Current experimental constraints on $|C_9|$ and $|C_{10}|$ are $\mathcal{O}(1)$\cite{Bause:2019vpr, Gisbert:2020vjx,LHCb:2022jaa}. For singlet VEVs $\sim 100 $ GeV, we see from Eqs.~\ref{eq:C9} and \ref{eq:C10} that in our model the Wilson coefficients are 1 to 2 orders of magnitude smaller than current constraints and hence leptonic $D$ meson decays are not expected to probe our model very effectively.

Considering the third generation couplings of the $Z$ in Eq.~\ref{eqZqq}, we note that while the $Z$ cannot decay into the top quark, the flavor changing $Z$ couplings to the top quark can induce  tree-level FCNC decays of the top quark~\cite{Aguilar-Saavedra:2004mfd, Alves_2024}, 
\begin{equation}
\Gamma_{t\rightarrow Z q}\simeq \frac{|g^u_{Z{\bar q}_Lt_L}|^{2}}{4\pi}\frac{m_{t}^{3}}{m_{Z}^{2}}\left(1-\frac{m_{Z}^{2}}{m_{t}^{2}}\right)^{2}\left(1+2\frac{m_{Z}^{2}}{m_{t}^{2}}  \right)\;.
\end{equation}

The branching ratios of $t\to Zu,~Zc$ are experimentally constrained to be $< 1.2\times10^{-4}$ \cite{ParticleDataGroup:2024cfk} and are strongly suppressed in the SM~\cite{Alves_2024} by the Glashow-Illiopoulos-Maiani (GIM) mechanism. Using the total top quark width,  $\Gamma = 1.42^{+0.19}_{-0.15}$ GeV \cite{ParticleDataGroup:2024cfk}, we find the constraint $g^u_{Z{\bar q}_L t_L} < 2.863\times10^{-2}$. This constraint allows values significantly larger than the couplings $g_{Z\bar{q}_Lt_L}^{u} \sim 10^{-5}$ in our model, obtained from Eq.~\ref{eqZqq} for singlet VEVs $\sim~100$~GeV. Hence corrections to $t\to Zu,~Zc$ are expected to be negligible in our model.

As a future target for electroweak precision experiments, assuming all the singlet VEVs are $\sim 150\text{~GeV}$, we write down the following coupling deviation matrix induced in our model for the $Z$ coupling to the left-handed up quarks, 
\begin{equation}\label{eqZ150}
\delta g_{Zq_L\bar{q}_L}^{u}\approx\begin{blockarray}{c c c c}
& u & c & t\\
\begin{block}{c@{\hspace{9pt}} [c c c]}
\bar{u}& 0.2 & 0.6 & -0.025  \\
\bar{c}&0.6 & 2.6 & -0.1\\
\bar{t}& -0.025 &-0.1 & 0.005  \\ 
\end{block}
\end{blockarray}~\times~ 10^{-4}\;\;.
\end{equation}

\subsection{Direct Searches at the LHC }
 The production of VLQs at the LHC mostly comes from gluon and photon mediated interactions. However, the VLQs can be both pair and singly produced by weak interactions mediated by the $W^{\pm},~Z,$ and $~H_t$. These interactions are limited due to the symmetries ($\mathbb{Z}_2^{Q_i}, \mathbb{Z}_2^{H_{{u}^j}}$) in our model and are mostly mediated by the longitudinal modes of the gauge bosons, i.e.~the Goldstones. Once produced, these VLQs can decay into up-type SM quarks and neutral bosons, which can again be either the Goldstone bosons ($G^0$), the SM Higgs ($H_t$), or the singlets; or down-type SM quarks and the charged Goldstone bosons ($G^\pm$). While the singlet masses are not constrained in our model, they are expected to be at the weak scale and hence should be kinematically accessible for the VLQ decays. As discussed earlier, other PHs in our model are significantly heavier than the SM Higgs. As such they are kinematically forbidden from appearing in final states for decays of VLQs that can be conceivably produced at the LHC.

Thus, considering the $\mathbb{Z}_2$ charges of our VLQs, only $\psi^{13}$ and $\psi^{23}$ decay via all the decay modes ($G^\pm:G^0:H_t:S$) = ($2:1:1:1$). On the other hand,  $\psi^{12}, \psi^{21}, \psi^{32},$ and $\psi^{31}$ decay dominantly through the singlet. Here, while the decay branching ratios to the goldstone modes and Higgs still go together, with a $(2:1:1)$ ratio, the branching ratios of $\psi^{ij}\rightarrow u^iH_t$ (where $j\neq 3$) are suppressed by $(v^j_u/v_t)^2$, driven by the small mixing of $H_u^j$ with $H_t$.   
Hence, the decay branching ratios for these are ($G^\pm : G^0 : H_t : S$) = ($(v^j_u/v_t)^2~\times\{2:1:1\}:1$). Table~\ref{tabDecay} summarizes the decays of our  VLQs with their corresponding branching ratios. 
\begin{table}[tbh]
	\begin{center}
		\begin{tabular}{|c|c|c|}
			\hline
			  	{}& $G^{\pm}:G^0:H_t$ & $S^{ij}$\\
			\hline
			SM quarks (q)& $2:1:1$ & $1$\\
			\hline
		    $u/d$& $\psi^{12} \left(\frac{v_c}{v_t}\right)^2, ~\psi^{13}$ & $\psi^{31},~ \psi^{21}$\\
		    \hline
		    $c/s$& $\psi^{21}\left(\frac{v_u}{v_t}\right)^2,~\psi^{23}$ & $\psi^{32},~\psi^{12}$\\
		    \hline
		    $t/b$& $\psi^{31}\left(\frac{v_u}{v_t}\right)^2,~\psi^{32}\left(\frac{v_c}{v_t}\right)^2$ & $ \psi^{13},~\psi^{23}$\\
		    \hline
		\end{tabular}
\caption{Decays of VLQs to SM quarks at the LHC with different branching ratios between $(G^{\pm}:G^0:H_t)$ and $S^{ij}$. In the second column, the factor $(v_u^j/v_t)^2$ next to $\psi^{ij}$ denotes the suppression of $\psi^{ij}\rightarrow d^i~G^{\pm}/u^iG^0/u^iH_t~(j\neq 3)$, as compared to  $\psi^{ij} \to u^j S^{ij}$ listed in the third column.
} \label{tabDecay}
\end{center}
\end{table}

Assuming $s_{ij} \sim 150$ GeV, which evades the $Z$ coupling, top decays, and FCNC constraints, from Eq.~\ref{eqMS}, we determine the approximate lower limit for our VLQ masses to be,
\begin{eqnarray}
M_{12} &\approx& M_{21} ~\sim~ 650 {\rm ~GeV}\,, \nonumber\\
M_{23} &\approx& M_{32} ~\sim~ 3.5 {\rm ~TeV}\,, \nonumber\\
M_{31} &\sim& 150 {\rm ~TeV}\,,\,M_{13} ~\sim~ 25 {\rm ~TeV}\;.
\label{eqmassre}
\end{eqnarray}
Hence, in our model, the VLQ masses are expected to be between $\sim 0.6$ and 150 TeV. The current LHC limits for direct detection of generic VLQs are between 1 and 2 TeV~\cite{Buckley:2020wzk, ATLAS:2024zlo}. Therefore, only the first generation of our VLQs, $\psi^{12}$ and $\psi^{21}$, conceivably lie within the current reach of the LHC. However, most of the LHC searches focus on VLQ partners of the top quark, i.e. final states with a top quark, but neither $\psi^{12}$ nor $\psi^{21}$ decay to a top quark. A recent search does consider decays into light quarks, but given the reduced branching ratios to $W^\pm,~Z,~H_t$ due to the presence of the singlets, there is currently no sensitivity~\cite{ATLAS:2024zlo,Banerjee:2024zvg}. We note that if the singlets are kinematically inaccessible for the VLQ decays, then the LHC excludes $\psi^{12}$ and $\psi^{21}$ masses $\sim 1100$~GeV~\cite{ATLAS:2024zlo}. Finally, for simplicity, we have assumed all Yukawa couplings to be $\mathcal{O}$(1), but small deviations may allow for a larger region of parameter space of our VLQ masses and couplings to be probed at the LHC.

\section{Summary}
To address the puzzles of the fermion mass hierarchy and the CKM flavor structure, we have proposed a novel BSM model that successfully resolves these problems. In our model, the hierarchical pattern observed in quark masses originates from a corresponding hierarchy in the Vacuum Expectation Values~(VEVs) of six different Higgs fields, the so-called Private Higgs~(PH) bosons. The VEVs of these  PH fields give masses to the quarks through $\mathcal{O}$(1) Yukawa couplings, making the quark masses and VEVs of the same order, thereby preserving the naturalness of the theory. The PH model is flavor diagonal by construction. Hence, to reproduce the CKM matrix, we introduce global $\mathbb{Z}_2$ symmetries for each quark generation and each  PH field, and a spectrum of singlet vector-like quark (VLQ) and singlet-scalar fields. The symmetries imposed constrain the possible interactions such that unwanted interactions~(such as tree-level FCNCs) can be eliminated.

We derive relationships between the up-type mass matrix,  CKM matrix, and physical quark masses. We also analyze the induced FCNC $Z$ interactions to understand the interplay between these elements in our model. Interestingly, we find that not only is the CKM matrix dictated completely by the structure of our model, but also it is {\it independent} of the fermion masses. These relationships enable us to constrain the masses of the VLQs and the singlet VEVs in our model and provide targets to probe it in future flavor and precision-electroweak experiments. 

Most of the PH bosons, as well as the second- and third-generation  VLQs, in our model are considerably heavy. Thus the direct impact of the LHC on our  model  is limited. 
While our lightest VLQs may be produced at the LHC, they decay only to the light quarks. Hence we motivate LHC searches for VLQs decaying exotically to light quarks and singlets rather than just the third generation and the SM gauge bosons and Higgs~(the vast majority of current LHC searches).   

Finally, we comment on the required CP-violating phase in the observed CKM. In our current setup, this phase would be generically induced due to the possible complex couplings and  VEVs of the PH and singlets in our model. These complex phases depend on the nature of the UV completion of our model and play no role in the fermion masses and mixing hierarchies constructed.  We leave investigations of  the origins of these complex phases and the scalar potential to future work.

In summary, our model provides an innovative framework for addressing key unresolved problems in particle physics. Through the incorporation of flavor symmetries, VLQ fields, and singlet scalars interacting with  PH bosons, we successfully reproduce the flavor structure of the SM and the CKM matrix, while preserving the naturalness of the theory. We provide motivation for future precision electroweak  and flavor experiments  as well as for LHC direct searches to probe our model.

\acknowledgments
We thank Carlos Wagner, Alexey Petrov, and Gil Paz for useful discussions. BB is supported by the U.S. National Science Foundation through Grant No. PHY-2310627. NRS is supported by U.S. Department of Energy under Contract No. DE-SC0007983. This work was performed in part at the Aspen Center for Physics, which is supported by National Science Foundation grant PHY-1607611. 

\bibliographystyle{JHEP}
\bibliography{Flavor_ref1}

\end{document}